SRDTI: Deep learning-based super-resolution for diffusion tensor MRI

Qiyuan Tian[1,2], Ziyu Li[3], Qiuyun Fan[1,2], Chanon Ngamsombat[1], Yuxin Hu[4], Congyu Liao[1,2], Fuyixue Wang[1,2], Kawin Setsompop[1,2], Jonathan R. Polimeni[1,2], Berkin Bilgic[1,2], Susie Y. Huang[1,2]

[1]Athinoula A. Martinos Center for Biomedical Imaging, Department of Radiology, Massachusetts General Hospital, Boston, MA, United States;

[2]Harvard Medical School, Boston, MA, United States;

[3]Department of Biomedical Engineering, Tsinghua University, Beijing, China;

[4]Department of Electrical Engineering, Stanford University, Stanford, CA, United States.

*Correspondence to: Qiyuan Tian, Ph.D., Athinoula A. Martinos Center for Biomedical Imaging, 149 13th Street, Charlestown, MA, 02129, United States. E-mail: qtian@mgh.harvard.edu.

**Synopsis**

High-resolution diffusion tensor imaging (DTI) is beneficial for probing tissue microstructure in fine neuroanatomical structures, but long scan times and limited signal-to-noise ratio pose significant barriers to acquiring DTI at sub-millimeter resolution. To address this challenge, we propose a deep learning-based super-resolution method entitled "SRDTI" to synthesize high-resolution diffusion-weighted images (DWIs) from low-resolution DWIs. SRDTI employs a deep convolutional neural network (CNN), residual learning and multi-contrast imaging, and generates high-quality results with rich textural details and microstructural information, which are more similar to high-resolution ground truth than those from trilinear and cubic spline interpolation.

**Introduction**

Diffusion tensor imaging (DTI) is widely used for mapping major white matter tracts and probing tissue microstructure in the brain[1,2]. High-resolution DTI (e.g., 1.25 mm isotropic adopted by the Human Connectome Project (HCP) WU-Minn-Oxford Consortium[3]), if it were widely available, would be beneficial for probing tissue microstructure in fine neuroanatomical structures, such as cortical anisotropy and fiber orientations[4]. However, long scan times and limited signal-to-noise ratio (SNR) pose significant barriers to acquiring DTI at high resolution in routine clinical and research applications. Higher resolutions would provide additional benefits[5-8], but it is challenging to acquire DTI data at millimeter and sub-millimeter isotropic resolution using standard 2D acquisitions with single-shot EPI readout, due to extremely high image blurring and distortion and low SNR. Slab acquisitions[5-9] and multi-shot EPI[10-12] have been employed for sub-millimeter DTI but require advanced sequences and reconstruction methods that are not widely available.

Super-resolution imaging provides a viable way to achieve DTI at higher resolution in the spirit of image quality transfer[13]. Previous studies have also demonstrated the feasibility of deep learning in super-resolution DTI[14]. To address this challenge, we propose a deep learning method entitled "SRDTI" to synthesize high-resolution diffusion-weighted images (DWIs) from low-resolution DWIs. Unlike previous studies using a shallow convolutional neural network (CNN), SRDTI employed a very deep 3D CNN, residual learning and multi-contrast information sharing. We compared our results to those from image interpolation and quantified the improvement in terms of both image similarity and DTI quality.

**Methods**

Data.

Pre-processed $T_1$-weighted (0.7-mm isotropic) and diffusion data (1.25-mm isotropic, b=1,000 s/mm$^2$, 90 uniform directions) of 200 healthy subjects (144 for training, 36 for validation, 20 for evaluation) from the HCP WU-Minn-Oxford Consortium were used[15]. Low-resolution data were simulated by down-sampling the high-resolution data to 2 mm iso. resolution using sinc interpolation. Co-registered $T_1$-weighted data were resampled to 1.25-mm isotropic resolution.

Network Implementation.

SRDTI utilizes a very deep 3D CNN[16-18] (10 layers, 192 kernels per layer) to learn the mapping from the input low-resolution image volumes to the residuals between the input and output high-resolution image volumes (residual learning) (Fig. 1). The inputs of SRDTI are a low-resolution b=0 image volume and six corresponding DWI volumes up-sampled to 1.25-mm isotropic resolution, and a $T_1$-weighted volume. The outputs of SRDTI are a ground-truth high-resolution b=0 image volume and six corresponding DWI volumes. The input b=0 image volumes were obtained by averaging all acquired b=0 image volumes. The DWI volumes were synthesized from the fitted diffusion tensor sampled along six optimized diffusion-encoding directions, which minimize the condition number of the diffusion tensor transformation matrix[19], and were therefore equivalent to the six diffusion tensor components in the image space. Operating in the image space rather than the tensor component space improved data similarity in local regions and avoids unreliable tensor fitting in cerebrospinal fluid voxels. Anatomical images are often acquired along with diffusion data and were therefore included as an additional channel to outline different tissues and preserve structural detail in the output DWIs. SRDTI was implemented using the Keras API (https://keras.io/) with a Tensorflow (https://www.tensorflow.org/) backend. Training was performed with 64×64×64 voxel blocks, Adam optimizer, L2 loss using a NVidia V100 GPU.

Evaluation.

For comparison, low-resolution diffusion data were also up-sampled to 1.25-mm isotropic using trilinear and cubic spline interpolation. The mean absolute error (MAE), peak SNR (PSNR) and structural similarity index (SSIM) were used to quantify image similarity comparing to ground-truth high-resolution images. The MAE of DTI metrics, including primary eigenvector (V1), fractional anisotropy (FA), mean, axial, and radial diffusivities (MD, AD, RD) and comparing to ground-truth high-resolution results were also calculated and compared.

**Results**

The b=0 image and DWIs at 1.25 mm isotropic resolution generated by SRDTI recovered more textural details and were visually similar to the ground-truth images (Fig. 2). The images were also quantitatively similar to the ground-truth images (Table 1a–c), with low MAEs around 0.012, high PSNRs around 31 dB and high SSIM around 0.98, which were 2 to 3 times better than those from trilinear and cubic spline interpolation. The residuals between the super-resolved images and ground-truth high-resolution images did not contain noticeable anatomical structure (Fig. 2, rows b, d, column iii).

The V1-encoded FA maps from SRDTI displayed captured the striated appearance of the gray matter bridges spanning the internal capsule in the striatum (Fig. 3i), as well as known cortical anisotropy (Fig. 3ii) and the fine fiber pathways in the pons (Fig. 3iii). Quantitatively, the MAEs of the output DTI metrics were also low, with MAEs of 8.41°±0.35° for V1, 0.022±0.0009 for FA, and 0.029±0.002 $\mu m^2$/ms, 0.038±0.0019 $\mu m^2$/ms and 0.03±0.0019 $\mu m^2$/ms for MD, AD and RD, which were 35% to 80% of the

MAE's calculated for the corresponding DTI metrics obtained from trilinear and cubic spline interpolated images (Table 1d).

**Discussion and Conclusion**

We obtained high-quality super-resolution DWIs and DTI metrics at 1.25 mm isotropic resolution from low-resolution DWIs at 2 mm isotropic resolution (~4.1× voxel volume difference) using SRDTI, which employs a very deep 3D CNN and residual learning. Our results recover detailed microstructural information and demonstrate substantial improvement over the results derived from trilinear and cubic spline interpolation. SRDTI can be generalized to high-b-value data for mapping crossing fibers and more advanced microstructural models (e.g., diffusion kurtosis imaging and NODDI[20]). SRDTI can also be used to super-resolve sub-millimeter isotropic resolution images obtained from slab acquisitions and multi-shot EPI as well as lower-resolution images acquired using standard 2D sequences with single-shot EPI. Future work will compare SRDTI to other super-resolution methods.

**Acknowledgments**

This work was supported by the NIH (grants P41-EB030006, U01-EB026996, R21-AG067562, K23-NS096056) and an MGH Claflin Distinguished Scholar Award.

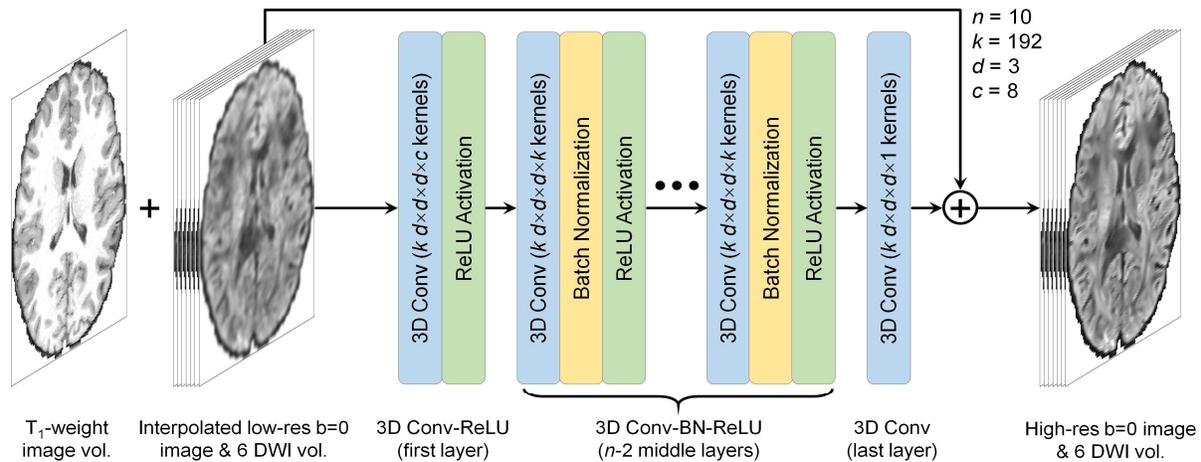

**Figure 1. CNN architecture.** A very deep 3-dimensional plain convolutional neural network (CNN) comprised of stacked convolutional filters paired with ReLU activation functions ($n$=10, $k$=192, $d$=3) is used for SRDTI. The CNN input is a b=0 image volume and six diffusion-weighted image (DWI) volumes along optimized diffusion-encoding directions interpolated to the target high resolution as well as an anatomical $T_1$-weighted image volume. The CNN output is the high-resolution b=0 image and six DWI volumes.

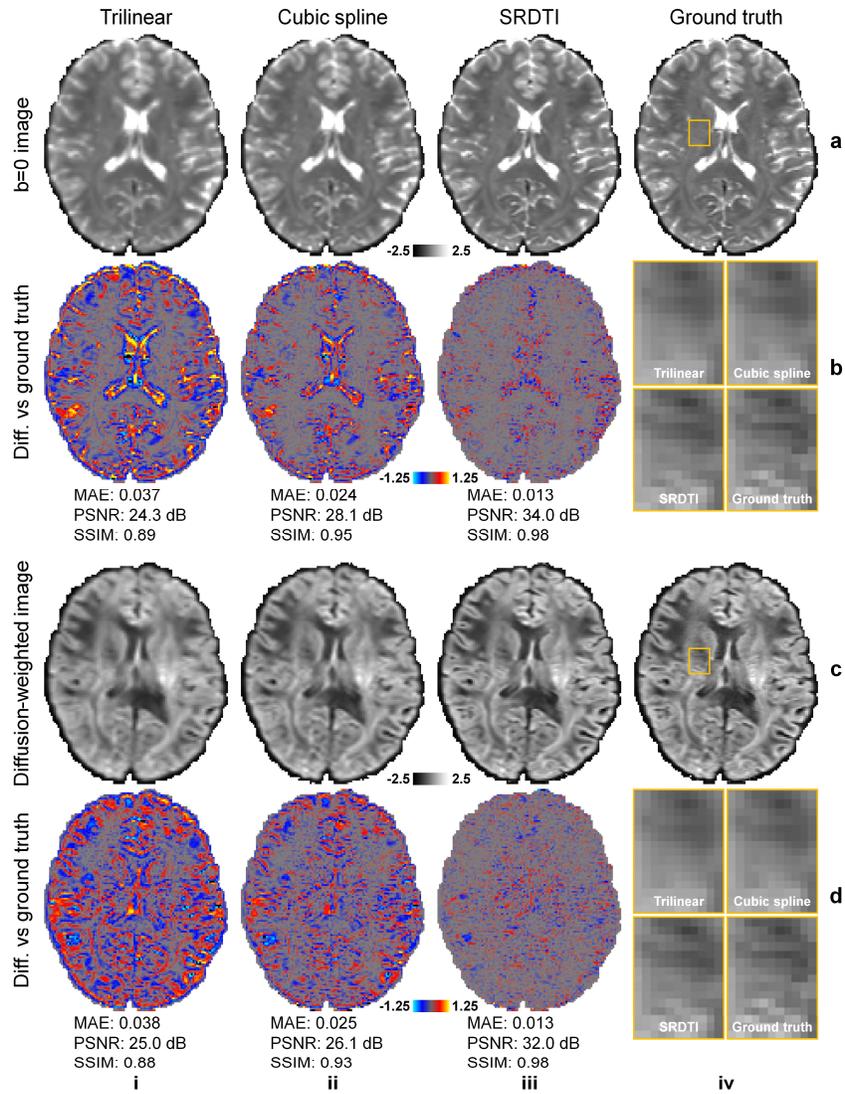

**Figure 2. Image results.** The b=0 images (row a) and diffusion-weighted images (DWIs) (row c) along one direction of the six optimized diffusion-encoding directions (i.e., [0.91, 0.416, 0]) from interpolated data (i, ii) (interpolated data using cubic spline is the input to SRDTI), SRDTI output data (iii), ground-truth high-resolution data (iv), and their residuals comparing to the ground-truth high-resolution images (rows b, d). A region-of-interest in the deep white matter (yellow boxes) is displayed in enlarged views (rows, b, d, column iv).

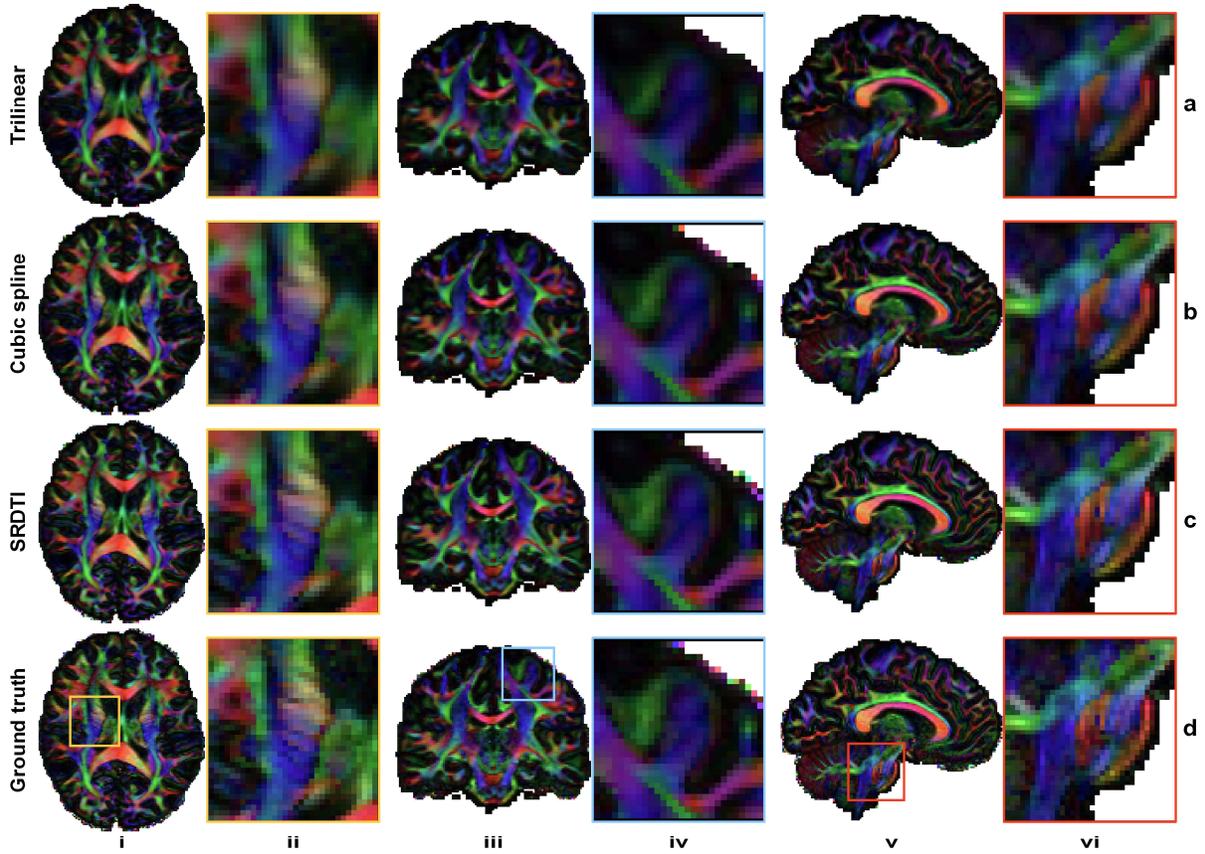

**Figure 3. Direction-encoded fractional anisotropy maps.** Fractional anisotropy maps color encoded by the primary eigenvector (red: left–right; green: anterior–posterior; blue: superior–inferior) derived from the diffusion tensors fitted using interpolated data (a, b), SRDTI super-resolution data (c), and ground-truth high-resolution data (d) along axial, coronal and sagittal directions, showing regions-of-interest in the internal capsule (i, ii), cerebral cortex (iii, iv) and pons (v, vi), respectively.

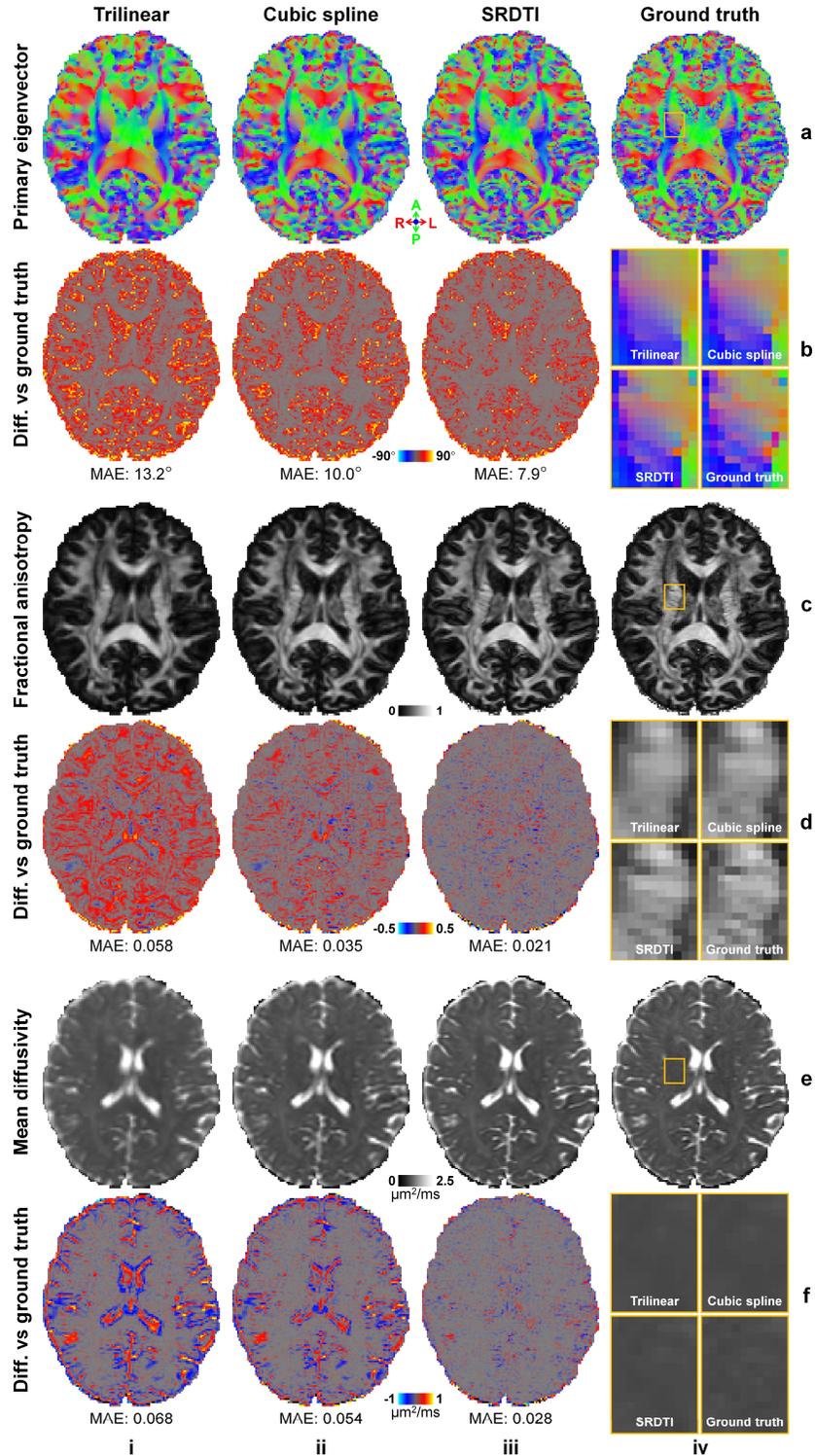

**Figure 4. Maps of DTI metrics.** Direction-encoded primary eigenvector (a), fractional anisotropy (c) and mean diffusivity (e) maps from interpolated data (i, ii), SRDTI super-resolution data (iii), ground-truth high-resolution data (iv), and their residuals comparing to the ground-truth high-resolution results (rows b, d). A region-of-interest in the deep white matter (yellow boxes) is displayed in enlarged views (rows, b, d, column iv).

### a                         Mean absolute error of images

|              | b=0 image      | DWI (dir.1)    | DWI (dir.2)    | DWI (dir.3)    | DWI (dir.4)    | DWI (dir.5)    | DWI (dir.6)    |
|--------------|----------------|----------------|----------------|----------------|----------------|----------------|----------------|
| Trilinear    | 0.035±0.0019   | 0.038±0.0016   | 0.036±0.0013   | 0.037±0.0015   | 0.038±0.0016   | 0.038±0.0014   | 0.037±0.0015   |
| Cubic spline | 0.022±0.0012   | 0.025±0.0010   | 0.024±0.0009   | 0.024±0.0009   | 0.025±0.0010   | 0.024±0.0009   | 0.024±0.0009   |
| SRDTI        | 0.012±0.0006   | 0.013±0.0004   | 0.013±0.0004   | 0.013±0.0005   | 0.013±0.0004   | 0.013±0.0004   | 0.013±0.0005   |

### b                         Peak signal-to-noise ratio of images (dB)

|              | b=0 image  | DWI (dir.1) | DWI (dir.2) | DWI (dir.3) | DWI (dir.4) | DWI (dir.5) | DWI (dir.6) |
|--------------|------------|-------------|-------------|-------------|-------------|-------------|-------------|
| Trilinear    | 24.7±0.48  | 24.5±0.42   | 25.0±0.43   | 24.7±0.42   | 24.5±0.40   | 24.7±0.40   | 24.8±0.42   |
| Cubic spline | 28.4±0.45  | 25.3±0.71   | 25.7±0.76   | 25.5±0.72   | 25.3±0.70   | 25.6±0.74   | 25.5±0.73   |
| SRDTI        | 34.0±0.47  | 30.7±0.89   | 30.9±0.93   | 30.8±0.90   | 30.7±0.88   | 30.9±0.93   | 30.8±0.91   |

### c                         Structural similarity index of images

|              | b=0 image     | DWI (dir.1)   | DWI (dir.2)   | DWI (dir.3)   | DWI (dir.4)   | DWI (dir.5)   | DWI (dir.6)   |
|--------------|---------------|---------------|---------------|---------------|---------------|---------------|---------------|
| Trilinear    | 0.90±0.0049   | 0.88±0.0056   | 0.89±0.0046   | 0.88±0.0058   | 0.88±0.0057   | 0.89±0.0049   | 0.88±0.0060   |
| Cubic spline | 0.95±0.0030   | 0.93±0.0039   | 0.94±0.0034   | 0.93±0.0041   | 0.93±0.0040   | 0.94±0.0035   | 0.93±0.0041   |
| SRDTI        | 0.98±0.0014   | 0.98±0.0019   | 0.98±0.0019   | 0.98±0.0022   | 0.98±0.0018   | 0.98±0.0018   | 0.98±0.0021   |

### d                         Mean absolute error of DTI metrics

|              | Primary eigenvector (°) | Fractional anisotropy | Mean diffusivity ($\mu m^2/ms$) | Axial diffusivity ($\mu m^2/ms$) | Radial diffusivity ($\mu m^2/ms$) |
|--------------|-------------------------|-----------------------|---------------------------------|----------------------------------|-----------------------------------|
| Trilinear    | 13.69±0.40              | 0.060±0.0020          | 0.066±0.0071                    | 0.092±0.0059                     | 0.075±0.0069                      |
| Cubic spline | 10.52±0.39              | 0.036±0.0013          | 0.052±0.0053                    | 0.068±0.0049                     | 0.055±0.0051                      |
| SRDTI        | 8.41±0.35               | 0.022±0.0009          | 0.029±0.0020                    | 0.038±0.0019                     | 0.03±0.0019                       |

**Table 1. Similarity quantification.** The mean absolute error (a), peak SNR (b) and structural similarity index (c) of b=0 images and diffusion-weighted images from interpolated data and SRDTI super-resolution data comparing to the ground-truth high-resolution data. The MAE of DTI metrics, including primary eigenvector, fractional anisotropy, mean, axial, and radial diffusivities from interpolated and super-resolution data comparing to ground-truth high-resolution results (d).